\documentclass[aps,twocolumn,prl]{revtex4}
\usepackage{graphicx,epsfig}
\usepackage{color}
\usepackage{dcolumn}
\usepackage{bm}
\usepackage{psfrag} 
\usepackage{float}
\usepackage{plain}
\usepackage{tabularx}
\usepackage[latin1]{inputenc}
\usepackage{amsmath}
\usepackage{verbatim} 
\begin{document}

\title{Surrejoinder to the Comment on:\\
``Thermostatistics of Overdamped Motion of Interacting Particles''\\
 by Y. Levin and R. Pakter.}

\author{J.~S. Andrade Jr.$^{1,3}$, G.~F.~T. da Silva$^{1}$, A.~A.
  Moreira$^{1}$, F.~D. Nobre$^{2,3}$, E.~M.~F. Curado$^{2,3}$}

\affiliation{$^{1}$Departamento de F\'{\i}sica, Universidade Federal
  do Cear\'a, 60451-970 Fortaleza, Cear\'a, Brazil\\ $^{2}$Centro
  Brasileiro de Pesquisas F\'{\i}sicas, Rua Xavier Sigaud 150,
  22290-180, Rio de Janeiro-RJ, Brazil\\$^{3}$National Institute of
  Science and Technology for Complex Systems, Rua Xavier Sigaud 150,
  22290-180, Rio de Janeiro-RJ, Brazil}

\maketitle

\section{Initial considerations}

The example investigated in Ref.~\cite{andradeprl10} corresponds to a
system of interacting vortices undergoing an overdamped motion, whose
dynamical behavior is studied through a molecular dynamics procedure
and the corresponding results, at the stationary state, are shown to
be in perfect agreement with the solution of a nonlinear Fokker-Planck
equation for $T=0$ as well as for finite values of $T$.

In their Rejoinder~\cite{levin2} Levin and Pakter repeat some of the points
raised in their previous Comment~\cite{levin1} (already refuted in our
first Reply~\cite{reply1}) and raise some new ones concerning our
recent publication~\cite{andradeprl10}. The present queries are
refuted in this Surrejoinder, whenever relevant for the results of
Ref.~\cite{andradeprl10}. It should be mentioned that in both
Comment~\cite{levin1} and Rejoinder~\cite{levin2}, one finds incorrect
statements, as well as a clear lack of knowledge about important
advances in the area related to Ref.~\cite{andradeprl10}. In what follows,
we highlight some misleading points found in Refs.~\cite{levin2,levin1}.

(a) In Ref.~\cite{levin1} the authors analyze the case $T=0$, and
treat the system by {\it replacing the interactions acting on a given
  particle by a potential}, showing that this potential satisfies an
inhomogeneous Helmholtz equation. By solving such an equation, the
authors claim to have obtained {\it an exact solution for this
  problem}. Clearly, this is a misleading statement, since the authors
have used {\it a mean-field type of approximation} to obtain the
differential equation that led to their solution. This point has
already been refuted, as can be seen on item (5) of our
Reply~\cite{reply1}, and we reinforce it herein. Unfortunately, the
authors can not understand this fundamental restriction of their
solution.

(b) In both Refs.~\cite{levin2} and \cite{levin1} by Levin and Pakter,
one can notice very clearly that these authors did not read
Ref.~\cite{andradeprl10} carefully.  Our theory, like all other
theories, has its validity subjected to some conditions. One of the
most important conditions for this theory concerns the fact that the
two-particles density may be approximated by the product of two
one-particle densities, i.e., $\rho^{(2)}(x, x\prime,t) \approx
\rho(x,t) \rho(x\prime,t)$. All counter-examples shown in
Refs.~\cite{levin1,levin2} violate this basic requirement, clearly
written in our Letter~\cite{andradeprl10}. It is not surprising that
the the results obtained by these authors are in contrast with ours.
In this Surrejoinder, we show an example of this, when we consider a
particular counter-example used in the Rejoinder by Levin and
Pakter~\cite{levin2}, namely, the potential $\exp(-x^4)$, and simulate
it satisfying our conditions: the result is exactly what was predicted
by our theory. We can only regret that such a hasty reading of our
paper has led to a waste of (important) time of many scientists. We
are not willing to spend more time in explaining again what is already
clearly written in our paper.

(c) The fact that the system is described by Newton's laws does not 
necessarily imply that one should have a Boltzmann-Gibbs 
statistical mechanics; the authors seem to ignore this very
basic point.

(d) In both Comments~\cite{levin1,levin2}, the authors show to be
unaware of the vast literature developed in the last 15 years,
concerning nonlinear Fokker-Planck equations and their relation to
generalized entropies (different from the Boltzmann-Gibbs one); many
of these references are cited in Ref.~\cite{andradeprl10}. In the
same way that the linear Fokker-Planck equation may be related to the
Boltzmann-Gibbs entropy, as shown in standard statistical-mechanics
textbooks, one can also show that nonlinear Fokker-Planck equations
may be related to generalized entropies. The authors also do not seem to
understand this important point.

\section{Two different physical situations}

Consider the problem of a fixed amount of fluid confined to a
flat-bottom cylindrical container. Assuming that the fluid is
incompressible, simple hydrostatics says that the potential energy
$U_h$ of such system is given by
\begin{equation}
\label{hidro}
U_h=\frac{\rho_h g}{2} \int{dx dy~z^2},
\end{equation}
where $\rho_h$ is the density of the fluid, $g$ is the gravity, and
$z(x,y)$ is the height of the column of fluid over a certain point.
If energy if dissipated, the system should evolve towards the minimum
potential energy, with a constant $z$ along the container. This seems
quite obvious, but a specialist in liquids may present good evidence
to contest all these observations. A simple look inside a narrow glass
container filled with some fluid shows us the presence of a meniscus.
Such specialist could even solve exactly the shape of the meniscus for
a particular liquid and conclude that Eq.~(\ref{hidro}) is wrong as well 
as all conclusions derived from it. However, a narrow container is an unfair
situation. In a wide container, Eq.~(\ref{hidro}) works perfectly at
any point, but very close to the edge of the container. The
specialist in liquids could also point out that his/her approach works
better as it solves $z$ at any point of the system, far or close to
the edge. However, this approach is more specific rather than more
general, since the shape of the meniscus is highly dependent on the
nature of the liquid and container, while Eq.~(\ref{hidro}) will work
efficiently to any container, as long as it is not so narrow, filled
with any incompressible fluid.

In this contribution we will show that the effects observed by Levin
and Pakter~\cite{levin2,levin1}, in the same way as a meniscus in a
confined fluid, are negligible effects that are amplified in highly
confined regimes. In the same way as the hydrostatic example mentioned
above, one can solve this system, taking care of all microscopic
details of the particular interaction, and find the whole profile,
bulk and edge. We choose, however, a more general approach to this
problem, that describes the correct way the particles distribute
themselves in the bulk system. As we show in our
Letter~\cite{andradeprl10}, our approach works for systems where the
particles spread over a region that is large, when compared with the
typical interaction range over the particles. Moreover, our approach
is more general than Levin and Pakter solution~\cite{levin1} as it is
NOT dependent on the particular interaction.

\section{Physical argumentation in favor of Tsallis entropy as a
  maximum for the system under investigation}

In our Letter~\cite{andradeprl10}, we have concluded that, under
certain conditions, systems of interacting over-damped particles
evolve to a state that maximizes Tsallis entropy with $\nu=2$.
Contrary to what Levin and Pakter believe, this conclusion is not that
surprising but can be demonstrated quite simply. The present
derivation is a little less rigorous than the one we introduce in
Ref.~\cite{andradeprl10}, but easier to comprehend, and may help us to
make our point.

Let us assume that one can obtain the potential energy of a system of
interacting particles $U_s$ in the form,
\begin{equation}
\label{up}
U_s=\int{dx dy~u(x,y)},
\end{equation}
where $u(x,y)$ is the density of potential energy. To determine $u$ we
will use the simplifying assumption that the density of particles is
continuous and constant. In this case we can identify
$u(x,y)=\rho(x,y) U_1(\rho)$, where $\rho$ is the number of particles
per unity of area, and $U_1(\rho)$ is the potential energy of a single
particle due to the interaction with its neighbors. Under the
assumption that $\rho$ is continuous and constant, we have,
\begin{equation}
\label{u1}
U_1=\rho\int{dr^2 U(r)}=a\rho,
\end{equation}
where $a\equiv\int{dr^2 U(r)}$, and $U(r)$ is a radially symmetrical
interacting potential.
Including the energy due to an external potential, $U_e$, we obtain,
\begin{equation}
\label{tsallis}
U_s=\int{dx dy (a \rho + U_e) \rho}.
\end{equation}
We now consider that energy is dissipated until the system evolves
towards the minimum energy. As we explain in our
Letter~\cite{andradeprl10}, the term in $\rho^2$ in
Eq.~(\ref{tsallis}) is identified with the negative of the Tsallis
entropy with index $\nu=2$. Therefore such system evolves towards the
maximum Tsallis entropy.

Clearly this method is valid only when some conditions are met.
Firstly, we assume that $\rho$ is a continuous function that varies
slowly within the interaction range of a particle. Precisely, in the
derivation presented in Ref.~\cite{andradeprl10}, we assume that,
within the interaction range of the particles, we have
$\rho(r)=\rho_0+\nabla\rho\cdot\mathbf{r}$, that is, we neglect terms
of second order in the variation of the density profile, and assume
that $\rho$ varies linearly within the interaction range. The
inclusion into Eq.~(\ref{tsallis}) of the external potential $U_e$
also imposes that it varies slowly with the interaction range.
Situations where such condition do not hold include, but are not
limited to, abrupt changes in the external potential.

As we explicitly stated in our Letter~\cite{andradeprl10}, our
approach is not adequate when $\rho$ in not locally homogeneous, for
instance, hard-core potentials or potentials with an attractive part
should induce local fluctuations that can not be considered in our
method. Anther condition where it fails is with long-range potentials,
where the interaction range is not defined. Also, if $\rho$ goes to
zero at some point of the system, one can not assume the linear
variation beyond this point, as a negative density has no meaning.
This indicates that effects observed close to this point are not
accounted for by our method.  However, these effects are microscopic,
in the sense that they vanish for distances a few units away of the
interaction range.

To object to our method, Levin and Pakter considered systems with
interaction range of some arbitrary length $\lambda$, and present
results for particles confined to a region of only a few units of
$\lambda$ (see the horizontal axis of Fig.~1 of their
Comment~\cite{levin1} and Fig.~2 of their Rejoinder~\cite{levin2}). As 
already mentioned, our method does not account for effects close to the 
edge of the profile, and since Levin and Pakter confine the particles to 
the extreme, these edge effects appear rather amplified in their results.
In our Reply~\cite{reply1}, we asked why Levin and Pakter decided to
present results for this extreme case rather than in the conditions we
considered and effectively studied (in their notation, this would be
achieved by simply making $q^2N=800$). In this condition, the
particles spread over a region more than a $100\lambda$ wide and edges
effects become negligible. In Fig.~1 here we used {\it their
analytical solution} to demonstrate that the effect they expect is
indeed negligible for a less confined system. We ask now Levin and
Pakter to stop diverging the subject. Do they admit that the blue
curve of Fig.~1 is their analytical solution?  Do they admit that any
deviation of our solution is an edge effect that becomes negligible
when the particles spread over several units of the interaction range
$\lambda$? If they do not admit these statements, what are their
evidences for the contrary? If they do admit it, why didn't they write
it clearly in their Comment~\cite{levin1}?

\section{Reply to Levin and Pakter's Rejoinder~\cite{levin2}}

1) Levin and Pakter claim that the thermodynamic limit can be obtained
only in the two ways: (a) by scaling the vortex charge to leave it
proportional to $\sqrt{N}$; and (b) by rescaling the confining
potential strength at the same time the number of particles grows,
$\alpha\sim N$.

This statement is nothing but false. What Levin and Pakter appear not
to know, but it is well explained in any introductory text on
statistical mechanics, is that one obtains the thermodynamic limit by
increasing the number of particles, while leaving the intensive
properties of the system unchanged. For instance, think of $N$
particles confined in a box of volume $V$. To obtain the thermodynamic
limit, one should scale both $V$ and $N$, keeping the average density
$N/V$ constant. The outlandish scaling proposed by Levin and Pakter
does not keep constant the density of particles that, in the
asymptotic limit of this scaling, diverges to infinity. Therefore,
although this scaling may result in a invariant form for $\rho/N$, it
is not the correct way to obtain the thermodynamic limit. To Levin and
Pakter benefit, we explain here how one should proceed to obtain the
thermodynamic limit in a system like ours. In the same way as in the
box mentioned previously, one should simply scale the dimension $L_y$
linearly with $N$, in order to keep $N/L_y$ constant. We invite Levin
and Pakter to verify that in this way, different from their scaling,
both the density profile and the average energy per particle do not
diverge, but remain invariant with $N$.

\vspace{0.25cm}

2) Levin and Parker state that a differential equation for the
particle density only makes sense in the limit where $N\to\infty$.

We agree. As we explained in item (1), it is possible to obtain this
limit in a simple and straightforward way. For some reason, this
escaped Levin and Pakter comprehension.

\vspace{0.25cm}

3) Levin and Pakter state that ``even allowing the incorrect
coarse-graining procedure of Andrade et al., (...) the density
distribution will not be given by a q-exponential'' since a
q-exponential is a smooth function and can not explain the
discontinuity observed in the edge of the density function. Moreover,
Levin and Pakter state that this discontinuity is observed in our
results.

If present, this discontinuity is surely not a relevant effect in the
regime investigated by us, that is, less confined systems where
particles spread over several units of the interaction range
$\lambda$. In our first Reply~\cite{reply1} to their
Comment~\cite{levin1}, we included in the inset of Fig.~1 (for
clarity, repeated in Fig.~1 of this Surrejoinder) the solution
proposed by Levin and Pakter for the conditions we studied. It became
already clear from these results how negligible, if present, this
effect would be in this regime. Regarding the numerical results
presented in their Rejoinder~\cite{levin2}, since they do not provide
details of their simulations, we can not fathom what Levin and Pakter
did wrong in their simulations, or even if they are really integrating
the correct equations of motion [see Eq.~(9) of our Letter].
Considering the evident asymmetry and non-smooth aspect of the curve
presented in Fig.~1 of their Rejoinder~\cite{levin2}, we can only
guess that they did not perform an average over enough realizations.
In any event, we now show in Fig.~1 of the present Surrejoinder our
numerical solution along with their numerical and analytical solutions
for the regime with $q^2N=800$, that reproduces the regime we
investigate. As one can see, although there is a good agreement
between our and their numerical solution, the analytical derivation
does not follow the observed profile. We can, however, use Levin and
Pakter's qualitative solution at this regime to demonstrate that the
discontinuity, if present, is negligible at the regime where the
particles are not so confined.

\begin{figure}[t]
\begin{center}
\includegraphics*[width=8cm]{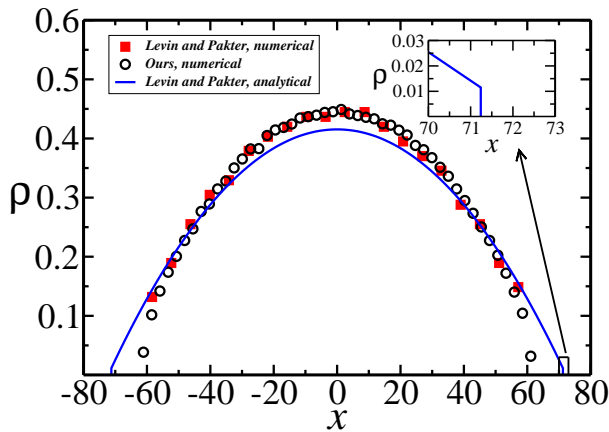}
\end{center}
\protect\caption{Numerical results for the density profile obtained
  from numerical simulations. We used the same conditions as Fig. 1 of
  our Letter~\cite{andradeprl10}, namely, $N=800$, $\alpha=10^{-3}
  f_{0}/\lambda$, and $L_{y}=20 \lambda$. The black circles and red
  squares correspond to the numerical simulation reported in
  Ref.~\cite{andradeprl10} and in the Rejoinder by Levin and 
  Pakter~\cite{levin2}, respectively, while the straight blue line corresponds 
  to the solution according to Eq.~(3) from Levin and Pakter's
  Comment~\cite{levin1}.  Clearly their solution is not compatible
  with the concavity of the density profile obtained from our
  molecular dynamics simulations. The inset shows that the
  discontinuity in their solution is barely noticeable for these
  physical conditions.}
\end{figure}

\vspace{0.25cm}

4) Levin and Pakter ask ``for exactly what value of $N$ do the authors
expect that Tsallis entropy will start determining the particle
distribution at T=0?''

If Levin and Pakter had read our Letter with attention, they would know
by now that as few as $800$ particles are enough to observe a
distribution that follows our predictions and maximizes Tsallis
entropy.

\vspace{0.25cm}

5) Levin and Pakter state that we do not provide ``any reason or
indication why they believe that the standard Boltzmann-Gibbs
statistical mechanics will not apply to the system studied by them.''

We do not question that other approaches using BG statistics could be
applied to solve the problem we have investigated. We are sure,
however, that if correctly applied, these approaches should
corroborate our results. Maybe Levin and Pakter still do not
understand our results. In our approach, all the particle-particle
interactions of the system are included into the Tsallis entropic term
with $\nu=2$. For the case with $T>0$, this means that we map the
problem of a non-ideal gas of particles into an ideal gas with an
entropy that caries both BG and Tsallis contributions. The usefulness
of our approach is that it can be generalized to a wide range of
interaction potentials. By contrast, in order to treat the same
problem only with BG statistics, one would need to include the
particular interaction, which requires the solution to be specific 
for each type of interaction potential.

\begin{figure}[t]
\begin{center}
\includegraphics*[width=8cm]{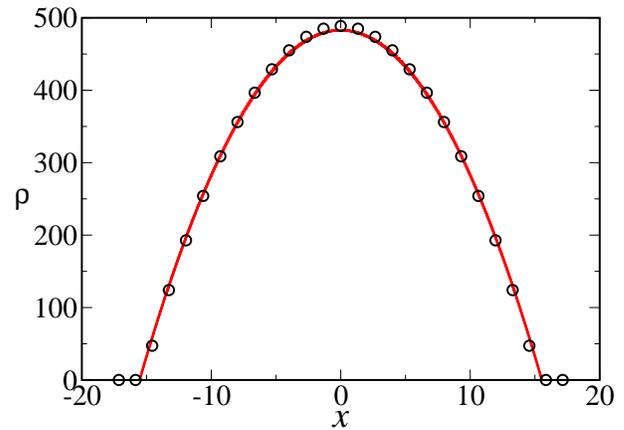}
\end{center}
\protect\caption{Numerical results for the density profile obtained
  from numerical simulations of a one-dimensional system and the
  interaction potential proposed in the Rejoinder by Levin and
  Pakter~\cite{levin2}, namely, $U(x)=f_{0}\exp{[-(x/\lambda)^{4}]}/2$.
  This potential approaches a square-barrier, being rather difficult
  to simulate through standard molecular dynamics techniques. In order
  to obtain minimum energy configurations, we coupled a
  steepest-descent Monte Carlo algorithm. The system has $N=10000$
  particles confined by an external potential with intensity
  $\alpha=3.0f_{0}/\lambda$. The solid line corresponds to the
  theoretical prediction Eq.~(14) of our Letter~\cite{andradeprl10}
  with $a=0.75$. As depicted, when the particles spread over several
  units of $\lambda$, we observe a parabolic profile. This is in
  perfect agreement with our theoretical model. }
\end{figure}

\vspace{0.25cm}

6) Levin and Pater state that from ``Eq.~(1) used by Andrade is
derived the unlucky Eq.~(13) of their PRL'', and that this equation
leads to the conclusion that the stationary state for particles
interacting by {\it any short-range force} is always parabolic. Levin
and Pakter also present numerical results indicating that this result
is not true for a potential given by $U(x)\sim e^{-x^4}$.

First, Levin and Pakter should note that we do NOT deduce Eq.~(13)
from Eq.~(1), but from the equations of motion for the particles,
Eq.~(9) of the Letter~\cite{andradeprl10}. Equation~(13), however, is
equivalent to the nonlinear Fokker-Plank equation, Eq.~(5), that, as
we show, drives the system towards the state that maximizes Tsallis
entropy. Second, we invite Levin and Pakter to investigate this system
in the conditions we know our method to be valid, that is, a system
with a less intense confining potential, where particles spread over
several units of the interaction range of the potential, instead of
only four. We believe that had they followed these guidelines, by now
they would know that this results in a parabolic density profile, as
shown in Fig.~2.  In short, as long as a few conditions are fulfilled,
a wide family of short-range repulsive interaction potentials can be
modeled by Eq.~(13) of our Letter~\cite{andradeprl10}.

\vspace{0.25cm}

7) Levin and Pakter claim that the velocity distribution follows
Maxwell-Boltzmann.

We did not study velocity distributions in Ref.~\cite{andradeprl10},
although this is an interesting property which deserves a careful
analysis on its own.  Herein, we notice again, as in many other parts
of their Comments~\cite{levin2,levin1}, that they state as ours,
assumptions that we have never made. Could they indicate any
allusions made in Ref.~\cite{andradeprl10} concerning velocity
distributions? This is not the best way to carry a fair scientific
discussion.

\vspace{0.25cm}

8) Levin and Pakter state that our system ``has nothing to do with
Tsallis statistics.''

We believe that we have demonstrated that the system do evolve to the
state of maximum Tsallis entropy. We are sorry that Levin and Pakter
neither present any evidence of the contrary nor believe in our
results.

\section{Final considerations}

It is our understanding that, in their two
Comments~\cite{levin2,levin1}, Levin and Pakter did not give any
relevant contributions to the problem addressed in
Ref.~\cite{andradeprl10}. Their arguments are, most of the times misleading. 
In our two replies, we have refuted their criticisms in full detail. We 
thus consider the present discussion as closed.

\end{document}